
%
\input harvmac
%
%
%
%
\ifx\answ\bigans
\else
\output={
  \almostshipout{\leftline{\vbox{\pagebody\makefootline}}}\advancepageno
}
\fi
%
%
%
\def\mayer{\vbox{\sl\centerline{Department of Physics 0319}%
\centerline{University of California, San Diego}
\centerline{9500 Gilman Drive}
\centerline{La Jolla, CA 92093-0319}}}
%
%

%
%
%
%
\def\abstract#1{\centerline{\bf Abstract}\nobreak\medskip\nobreak\par #1}
%
%
%
%
\edef\tfontsize{ scaled\magstep3}
 \tfontsize  \tfontsize
 \tfontsize \font\titlei=cmmi10 \tfontsize
\font\titleis=cmmi7 \tfontsize \font\titleiss=cmmi5 \tfontsize
\font\titlesy=cmsy10 \tfontsize \font\titlesys=cmsy7 \tfontsize
\font\titlesyss=cmsy5 \tfontsize  \tfontsize
\skewchar\titlei='177 \skewchar\titleis='177 \skewchar\titleiss='177
\skewchar\titlesy='60 \skewchar\titlesys='60 \skewchar\titlesyss='60
%
%
%
%
%
\def\inv{^{\raise.15ex\hbox{${\scriptscriptstyle -}$}\kern-.05em 1}}
\def\lbar{{\lower.35ex\hbox{$\mathchar'26$}\mkern-10mu\lambda}} 

%
%
%
%
\def\dsl{\,\raise.15ex\hbox{/}\mkern-13.5mu D} 
\def\delsl{\raise.15ex\hbox{/}\kern-.57em\partial}
\def\Ksl{\hbox{/\kern-.6000em\rm K}}
\def\Asl{\hbox{/\kern-.6500em \rm A}}
\def\Dsl{\hbox{/\kern-.6000em\rm D}} 
\def\Qsl{\hbox{/\kern-.6000em\rm Q}}
\def\gradsl{\hbox{/\kern-.6500em$\nabla$}}
%
%
\def\lspace{\ifx\answ\bigans{}\else\qquad\fi}
\def\lbspace{\ifx\answ\bigans{}\else\hskip-.2in\fi} 
%
%
\def\boxeqn#1{\vcenter{\vbox{\hrule\hbox{\vrule\kern3pt\vbox{\kern3pt
        \hbox{${\displaystyle #1}$}\kern3pt}\kern3pt\vrule}\hrule}}}
%
%
\def\mbox#1#2{\vcenter{\hrule \hbox{\vrule height#2in
\kern#1in \vrule} \hrule}}
%
%
%
%

%
%
%
%
%

%

\def\bar#1{\overline{#1}}

\def\darr#1{\raise1.5ex\hbox{$\leftrightarrow$}\mkern-16.5mu #1}

%
%
\def\frac#1#2{{\textstyle{#1\over #2}}} 
%
%
%
%

\def\Im{\mathop{\rm Im}}
\def\Re{\mathop{\rm Re}}

%
%
%
%

%
%
\def\ltap{\ \raise.3ex\hbox{$<$\kern-.75em\lower1ex\hbox{$\sim$}}\ }
\def\gtap{\ \raise.3ex\hbox{$>$\kern-.75em\lower1ex\hbox{$\sim$}}\ }
\def\gl{\ \raise.5ex\hbox{$>$}\kern-.8em\lower.5ex\hbox{$<$}\ }
\def\roughly#1{\raise.3ex\hbox{$#1$\kern-.75em\lower1ex\hbox{$\sim$}}}
%
%

%

%

\relax

\def\dpart{\partial\kern .5ex\llap{\raise
1.7ex\hbox{$\leftrightarrow$}}\kern -.7ex {_\mu}}

\def\frac#1#2{{\textstyle{#1 \over #2}}}

\def\piee{\pi^0\rightarrow e^+e^-}
\def\etaee{\eta\rightarrow e^+e^-}
\def\etamumu{\eta\rightarrow \mu^+\mu^-}
\def\Im{{\rm Im}}
\def\Re{{\rm Re}}
\def\Br{{\rm Br}}

\def\s2weak{\sin^2\theta_{\rm w}}

\def\({\left(}\def\){\right)}

\def\mayer{\vbox{\sl\centerline{Department of Physics}
\centerline{9500 Gilman Drive 0319}
\centerline{University of California, San Diego}
\centerline{La Jolla, CA 92093-0319}}}
\def\caltech{\vbox{\sl\centerline{California Institute of Technology}
\centerline{Pasadena, California 91125}}}

\def\[{\left[}
\def\]{\right]}
\def\({\left(}
\def\){\right)}

\noblackbox
\vskip 1.in
\centerline{{\titlefont{The Rare Decays  $\piee$, $\etaee$ and }}}
\medskip
\centerline{{\titlefont{$\etamumu$ in Chiral Perturbation Theory}}}
\bigskip
\vskip .5in
\centerline{
Martin J. Savage\footnote{$^{\dagger}$}
{SSC Fellow}
and Michael Luke}
\bigskip
\mayer
\bigskip
\vskip .2in
\centerline{Mark B. Wise}
\bigskip
\caltech
\bigskip\bigskip
\centerline{{\bf Abstract}}
\bigskip
We calculate the decay rates for $\piee$, $\etaee$ and $\etamumu$
in chiral perturbation theory.
The linear combination of counterterms
necessary to render these amplitudes finite is fixed
by the recently measured branching fraction for $\etamumu$.
We find $\Br(\piee ) = 7\pm 1\times 10^{-8}$ and
$\Br(\etaee )=5\pm 1\times 10^{-9}$.

\vfill
\hbox{\hbox{ UCSD/PTH 92-23\hskip 2in July 1992} }
\hbox{\hbox{ CALT-68-1803} }
\eject

\newsec{Introduction}

In this letter we use chiral perturbation theory to calculate the
electromagnetic decays of the $\pi^0$ and $\eta$ to $\ell^+\ell^-$.
These decays proceed through two photon intermediate
states containing the anomalous $\pi^0\gamma\gamma$ $(\eta\gamma\gamma)$
coupling as
well as from a local $\pi^0\ell^+\ell^-$ $(\eta\ell^+\ell^-)$ operator
which is
required as a counterterm to render the one-loop diagram finite.  We
determine the coefficient of the counterterm by fitting to a recent
measurement of $\Br(\etamumu)$ \ref\sat{SATURNE measurement, reported in
CERN Courier, May 1992, p.10.}, which then allows us to predict the
rates for $\etaee$ and $\piee$.

Our work differs from previous calculations in which
a hadronic form factor is associated with the $\pi^0\gamma\gamma$
and $\eta\gamma\gamma$ vertices
\ref\geff{D.A. Geffen and B.L. Young, Phys. Rev. Lett.,
{\bf 15} (1965) 316.}\nref\quigg{C. Quigg and J.D. Jackson, UCRL-18487
(1968), unpublished.}\nref\babu{K.S. Babu and E. Ma, Phys. Lett.
{\bf 119B} (1982) 449.}\nref\kamal{A.N. Kamal and L. Chong-Huah,
Phys. Rev. {\bf D32} (1985) 1744.}\nref\berga{L. Bergstrom, Z. Phys.
{\bf C14} (1982) 129.}\nref\bergb{L. Bergstrom {\it et al}, Phys. Lett.
{\bf B126} (1983) 117.}-\ref\land{L.G. Landsberg, Phys. Rep.
{\bf 128} (1985) 301.}.
This makes the one loop integral finite but introduces model dependence
into the dispersive piece of amplitude.  The absorptive piece of the
amplitude is related to the $\pi^0 (\eta)\rightarrow \gamma\gamma$ width
by unitarity and is therefore unambiguous.

A precise theoretical prediction for these decays is interesting not
only in itself, but also because the $\eta\mu\mu$ and $\pi^0\mu\mu$
couplings contribute to the ``background'' for parity
violating observables
in $K^+\rightarrow\pi^+\mu^+\mu^-$, as has been emphasized recently in
refs.~\ref\sw{M.~J.~Savage and M.~B.~Wise, Phys. Lett {\bf B250} (1990)
151.}\ and \ref\lsw{M.~Lu, M.~B.~Wise and M.~J.~Savage, UCSD and Caltech
preprint
UCSD/PTH 92-20, CALT-68-1798 (1992).}.  These couplings also provide a
``background'' to the $T$ odd observables in this decay which have been
investigated in refs.~\sw\ref\triumf{P. Agrawal {\it et al}, Phys. Rev. Lett.
{\bf 67} (1991) 537; P. Agrawal {\it et al}, Phys. Rev. {\bf D45} (1992)
2383.}.
The $\mu^+$ spin polarisation is a
parity violating
observable whose magnitude gets a contribution from short distance
physics.  In the standard model this is dominated by top quark loops
and so provides a measurement of the real part of the
CKM matrix element $V_{td}$ in the phase convention where $V_{\rm bc}$
is real.
The long-distance physics  background from the $\eta\mu\mu$ and
$\pi^0\mu\mu$
couplings has been studied in detail in \lsw\ where it was found
to be significant for small values of $\Re V_{\rm td}$ and small top
quark masses.
Therefore, it is important to understand these electromagnetic rare
decays of the $\pi^0$ and $\eta$ in order to form a reliable estimate of
the background to the  determination of $Re V_{\rm td}$ from
$\mu^+$ polarisation measurements.

\newsec{$\etamumu$}

The graphs in \fig\graphs{Leading graphs contributing to $\etamumu$.}
give the leading contribution to $\etamumu$ in
chiral perturbation theory.
The $\eta\gamma\gamma$ and $\pi^0\gamma\gamma$ vertices arise from the
Wess-Zumino term
\ref\wess{J. Wess and B. Zumino, Phys. Lett., {\bf B37} (1971) 95.}
\eqn\wz{{\cal L}_{\rm WZ} = {\alpha\over 4\pi f}
\epsilon_{\mu\nu\lambda\sigma}F^{\mu\nu}F^{\lambda\sigma}\left(\pi^0/\sqrt{2}
+ \eta/\sqrt{6}\right) + .......\ \ \ ,}
where $f=135$ MeV.  This leads to a decay width
\eqn\egam{\Gamma (\eta\rightarrow\gamma\gamma) =
{\alpha^2m_\eta^3\over 96\pi^3 f^2}\ \ \ .}
The imaginary part of the one-loop graph is finite and related by
unitarity to the width \egam \geff\quigg; however, the real part
diverges and requires a local counterterm
\eqn\loc{\eqalign{
{\cal L}_{\rm c.t.}
= {3i\alpha^2\over 32\pi^2}&
\overline{\ell}\gamma^\mu\gamma_5 \ell\
\left[\chi_1Tr(Q^2\Sigma^\dagger\partial_\mu \Sigma -
Q^2\partial_\mu\Sigma^\dagger\Sigma)\right.\cr
&\left. + \chi_2Tr(Q\Sigma^\dagger Q\partial_\mu\Sigma
-Q\partial_\mu\Sigma^\dagger Q\Sigma)\right]\ ,}}
where $\ell=e$ or $\mu$, $Q$ is the electromagnetic charge matrix
\eqn\em{Q=\left( \matrix{2/3&0&0\cr 0&-1/3&0\cr 0&0&-1/3\cr}
\right) \ \ .}
Each term in \loc\ contains two factors of $Q$ because
${\cal L}_{\rm c.t.} $ arises from Feynman diagrams where two photons
produce the $l^+l^-$ pair.
The field $\Sigma=\exp(i2M/f)$ is the usual exponentiation of the
goldstone boson matrix where
\eqn\gold{M = \left(\matrix{
\pi^0/\sqrt{2}+\eta/\sqrt{6}&\pi^+&K^+\cr
\pi^-&-\pi^0/\sqrt{2}+\eta/\sqrt{6}&\overline{K^0}\cr
K^-&K^0&-2\eta/\sqrt{6}\cr}\right)\ \ \ .}
The coefficients $\chi_1$ and $\chi_2$
are renormalisation scheme dependent and subtraction scheme dependent;
we use  dimensional regularisation with $\bar{MS}$  (the gamma
matrix algebra is performed in 4 dimensions) and
choose the subtraction point to be  $\Lambda = 1$ GeV.
We find the width for $\etamumu$ to be
\eqn\etamm{\Gamma (\etamumu) = {\alpha^2 m_\mu^2 m_\eta\over 48\pi
f^2}|A(m_\eta^2)|^2\sqrt{1-{4m_\mu^2\over m_\eta^2}}\ \ \ ,}
where
\eqn\ima{ \Im A(s) = {\alpha\over\pi}{1\over\sqrt{1-\xi^{-2}}}
\log\left({1+\sqrt{1-\xi^{-2}}\over \xi^{-1}}\right)\ \ \ ,}
and \lsw
\eqn\rea{\eqalign{
\Re A(s) = {\alpha\over 4 \pi^2}&
\left[\chi_1(\Lambda)
+\chi_2(\Lambda) + 11
- 6\log\left({m_\mu^2\over\Lambda^2}\right)\right.\cr
&\left. + 2\xi^2 - 4\xi^4
+ 4\xi^2\log(4\xi^2) + 8\xi^4\log(4\xi^2)\right.\cr
&\left. -4\int_0^1dx \left[ 3+
{2(\xi^2-1)\sqrt{x}\over\sqrt{x+\xi^{-2}(1-x)}} \right]
\lambda_+^2\log|\lambda_+|\right.\cr
&\left. -4\int_0^1dx \left[ 3-
{2(\xi^2-1)\sqrt{x}\over\sqrt{x+\xi^{-2}(1-x)}} \right]
\lambda_-^2\log|\lambda_-|\right].}}
We have defined
$\xi^2 = s/4m_\mu^2$ and
$\lambda_\pm = \sqrt{x\xi^2}\pm\sqrt{x\xi^2 + (1-x)}$.
This amplitude is renormalisation scheme independent.
A change in the value of $\log\left({m_\mu^2\over\Lambda^2}\right)$
due to a different choice of
$\Lambda$ is compensated by a change in
the coefficient $\chi_1(\Lambda)+\chi_2(\Lambda)$.

The real part of the amplitude agrees with previous computations which
introduced a form factor for the $\eta\gamma\gamma$ vertex \geff
--\land\ when we take the mass associated with the form factor to be
large and retain only the leading term.

The branching fraction for $\etamumu$ has recently beeen remeasured at
SATURNE \sat, a machine dedicated to $\eta$ physics, which finds
$\Br(\etamumu ) = (5\pm1)\times 10^{-6}$, near
the unitary limit of $4.3\times 10^{-6}$ set by
$\Im A(m_\eta^2)$.
This fixes the sum of the counterterms
$-40 < \chi_1(\Lambda)+\chi_2(\Lambda) < -13 $.
Note that the rate is relatively insensitive to the precise value of
the counterterms.  This is because the one loop amplitude is infrared
divergent as $m_\ell\rightarrow 0$ and so dominates the contribution
from the counterterm.  Similarly, in phenomenological models it has been
found that the predictions for the branching ratio are relatively
insensitive to the exact form and scale of the hadronic form factors.

\newsec{Decays to $e^+e^-$}

Having fixed the sum of counterterms
$\chi_1(\Lambda)+\chi_2(\Lambda)$
we may now unambiguously predict the rates for
$\etaee$ and $\piee$.
It is important that $\chi_1$ and $\chi_2$ are the same for the cases
$l=e$ and $l=\mu$.  This occurs because both the $e$ and $\mu$ masses
are small compared with the chiral symmetry breaking scale.
{}From expressions analogous to \etamm --\rea\
found by substituting $m_e$ for $m_\mu$
and evaluating at either $s=m_\eta^2$ or $s=m_\pi^2$
(in the $\pi^0$ case \egam\  and \etamm\  are multiplied by 3)
we find
\eqn\eebr{\eqalign{&\Br(\piee)=7\pm 1\times 10^{-8}\cr
&\Br(\etaee)=5\pm 1\times 10^{-9}}}
compared to the present experimental upper bounds \ref\pdb{Particle Data
Group, Phys.~Rev.~{\bf D45} (1992) S1.}
\eqn\expt{\eqalign{&\Br(\piee)_{\rm exp.}<1.3\times 10^{-7}\cr
&\Br(\etaee)_{\rm exp.}<3\times 10^{-4}.}}

For $\piee$, the present upper limit is within a factor of two of the
theoretical prediction, and one may hope that in the near future this
decay mode will be observed.
A precise determination of the branching ratio
would test the validity of chiral perturbation theory for these decays.
In contrast, the experimental upper limit for $\etaee$ is five orders of
magnitude above the theoretical prediction.  This
upper limit was determined in a bubble chamber experiment
performed in 1966 with $\sim 10^4$ $\eta$'s\ref\bubble{M.J. Esten {\it
et al}, Phys.~Rev.~{\bf B24} (1967) 115.};  hopefully this limit will
be dramatically improved at SATURNE where $10^8$ $\eta$'s are produced
per day.

Corrections to our predictions come from higher
dimension operators in the chiral Lagrangian which contain more
derivatives or more factors of $m_s$.
These are suppressed by factors of $m_\eta^2/\Lambda_\chi^2\sim$ 25\%.

\bigskip

We acknowledge the support of U.S. Department of Energy under Contracts
DEAC-03-81ER40050 and DE-FG03-90ER40548, and NSF grant PHY-9057135.
MJS acknowledges the support of a Superconducting Supercollider National
Fellowship from the Texas National Research Laboratory Comission
under grant  FCFY9219.

\listrefs
\listfigs
\end